\documentclass[10pt, secnumarabic, nobibnotes, amssymb,notitlepage,pra,aps,twocolumn]{revtex4-1}
\usepackage{amsmath}
\usepackage{amsfonts}
\usepackage{amssymb}
\usepackage{hyperref}
\usepackage{graphicx}
\usepackage{color}
\usepackage{xcolor}
\usepackage[mathscr]{euscript}
\usepackage{longtable}
\usepackage{rotating}

\newcommand{\ket}[1]{\left| #1\right\rangle}

\begin{document}

\title{Dynamics of a three-photon state interacting with a quantum dot}

\author{J.C. L\'opez Carre\~no$^{1,2\ast}$}

\author{J.P. Restrepo Cuartas$^{1}$}

\author{H. Vinck Posada$^{2}$}
\affiliation{$^{1}$Departamento de F\'isica Te\'orica de la Materia Condensada, Universidad Aut\'onoma de Madrid, 28049 Madrid, Spain\\
$^{2}$Departamento de F\'isica, Universidad Nacional de Colombia, Bogot\'a, Colombia
}

%
\begin{abstract}

In this paper, we study the dynamics of the interaction of a three-photon state and a quantum dot embedded in a semiconductor cavity. In the first place, we  consider an ideal cavity in which the effects due to the environment are neglected. Under this conditions, the most important feature of the dynamics is its periodicity, which may be seen on the temporal evolution of the light state. The entanglement dynamics, which is studied through the negativity and the linear entropy, shows a periodic behavior too.
On the other hand, when considering the interaction between the cavity and the environment, taking into account both an incoherent pumping and a photon leakage, the dynamics is no longer periodic. Instead, it may be seen that the entanglement reach stationary values, which depend on the incoherent pumping and photon leakage rates. Finally, we discuss the possibility of using this sort of systems for quantum information processing.

\end{abstract}

\maketitle

\section{Introduction}\label{sec:01}

The light-matter interaction in quantum dots embedded in semiconductor microcavities is an area that has been studied widely in the last few years \cite{Tejedor-2004, Vinck-2005, Yamamoto-2000, Haroche-2007}. Moreover, the experimental realization of this kind of physical system has brought new phenomenology, such as emitters of single photons or bundles of them \cite{Bundler-2014}, and sources of entangled quantum states \cite{Yamamoto-2002, Yamamoto-2003, Pelton-2003}. 

On the other hand, the generation of Fock states containing $n$ photons has been one of the most interesting research areas, since its applications are useful for quantum communication and quantum computation \cite{Kiesel-2003}. In particular, generation of  Fock states containing three photons may be obtained by means of a process known as \textit{ spontaneous parametric down conversion (SPDC)} \cite{Douady-2004, Corona-2011, Gravier-2008, Richard-2011, Dot-2012}, or by means of optic nonlinearities both in self assembled \cite{Persson-2004, Antonosyan-2011, Rodrigo-2011, Lopez-2014} and in organic systems \cite{May-2005}.

Previously, we have shown that a three photon state may be prepared in a semiconductor microcavity \cite{Lopez-2014}. As a consequence, in this paper we study the dynamics of a three-photon state interacting with a quantum dot inside a semiconductor microcavity, focusing on two particular cases: (a) an ideal cavity, in which we neglect all the incoherent processes that could perturb the system, and (b) a real cavity, in which we study the effects of the environment over the three-photon state.
To characterized the state of the system, we study the entanglement between the three photon state and the quantum dot as a function of time, by means of both the negativity and linear entropy. Moreover, we study the state of light using the Wigner representation, which to some extent gives us a qualitatively information of the state's density operator.

The rest of the article is organized as follows: in section \ref{sec:02}, we describe the theoretical model to describe and characterize the quantum state as well as its dynamics. Then, in the section \ref{sec:03} we justify the range of parameters used to compute the results and proceed to present and discuss the results. Finally, in the section  \ref{sec:04} we provide an overview of the results and conclude.

\section{Theoretical Model}\label{sec:02}

The system under consideration is composed by a quantum dot interacting with a three photon state of light inside a semiconductor microcavity. In the theoretical model used, we consider that most of the the electronic levels of the quantum dot are off-resonant with the electromagnetic mode, so the energy structure of the quantum dot may be reduced to two levels: its ground state and an excited state. Moreover, the three photon state is introduced into the cavity by processes of incoherent excitation pumping as well as incoherent photon leakage.

The state considered in this paper is a quantum state made of the superposition of Fock states containing $3n$ photons. In general, this sort of states may be written in the following way:
\begin{equation}\label{Estella-general}
\ket{*}=c_0 \ket{0} + c_1 \ket{3} + \cdots + c_{n} \ket{3n}.
\end{equation}
Nevertheless, we constraint our study to the first two terms in the previous equation (\ref{Estella-general}):
\begin{equation}\label{Estella}
\ket{*}=\beta \ket{0} + \sqrt{1-|\beta |^2}\ket{3},
\end{equation}
where $\beta$ is a free parameter characterizing the state, and its a complex number. The amplitude and phase of the parameter are associated to the shape and orientation of the Wigner representation of the state in the phase space.

\subsection{Description of the Hamiltonian}

The interaction between a mode of the electromagnetic field and a two-level system is described by the Jaynes-Cummings model \cite{JCModel}.  The Hamiltonian used to described our system is (we take $\hbar=1$ along the paper):
\begin{equation}\label{Hamiltoniano}
H=\omega_a a^{\dagger} a+\omega_{\sigma} \sigma^{\dagger}\sigma + g \left( a^{\dagger}\sigma+a \sigma^{\dagger} \right).
\end{equation}
Here $\omega_a$ and $\omega_{\sigma}$ are the electromagnetic mode frequency and transition frequency of the two-level system, respectively. 
Furthermore, we have introduce the ladder operators for both the two-level system ($\sigma$ and $\sigma^{\dagger}$) and the electromagnetic field ($a$ and $a^{\dagger}$). The former follows the Fermi statistics and describes the transitions between the base and excited state of the two-level system, whereas the latter describes the usual boson creation and annihilation processes.
At last, the third term in eq. (\ref{Hamiltoniano}) describes the interaction between the quantum dot and the electromagnetic field, where $g$ is the dipole coupling constant.

\subsection{Master equation}

The temporal evolution of the system is described by the master equation in the Lindblad form:
\begin{multline}\label{Ec. Maestra}
\dot{\rho}=i \left[\rho , H\right]+\frac{P}{2}\left(2\sigma^{\dagger}\rho\sigma -\sigma\sigma^{\dagger}\rho -\rho\sigma\sigma^{\dagger}\right)\\ + \frac{\kappa}{2}\left(2a \rho a^{\dagger} -a^{\dagger}a\rho -\rho a^{\dagger}a\right),
\end{multline}
Here, the first term correspond to the dynamics in an ideal cavity (without incoherent processes, or losses of any kind), and $H$ is the Hamiltonian given in eq.  (\ref{Hamiltoniano}). The second term describes an off-resonant pumping of excitation, and the last term takes into account the cavity losses.
The parameters $P$ and $\kappa$ are the rate at which the two-level system has a transition from its base state towards its excited state, and the rate at which the cavity losses one photon, respectively.

\subsection{Measurement of the entanglement}

We study the entanglement between the three-photon state and the quantum dot using the \textit{negativity} and the \textit{linear entropy}. The former was proposed in \cite{Negat}, and is based on the Peres' criteria \cite{Peres-1996}, which proved the required conditions that any density operator should satisfy in order to be separable. Later on, Horodecki {\it et al.}  showed that the Peres' criteria was a sufficient condition for separability in systems whose Hilbert space has dimensions either $2\otimes2$ or $2\otimes3$ \cite{Horodecki-1996}. Otherwise, the negativity fails to quantify the entanglement, but allows us to witness it (the negativity tells us that the system is entangled, but does not tells us \textit{how much} entanglement there is). 

The negativity is computed in the following way,
\begin{equation}\label{negatividad-def}
N\left (\rho\right )\equiv\frac{\Vert \rho^{T_1}\Vert-1}{2},
\end{equation}
where $\Vert \rho^{T_1}\Vert$ is the trace norm of the density operator $\rho^{T_1}$, which is the result of the partial transposition of the subsystem 1 in the density operator $\rho$.

On the other hand, the linear entropy has been studied in systems composed of two qubits, and has proven to be a good entanglement quantifier \cite{Faruya-1998, Bose-2000, Munro-2001}. Nevertheless, when considering dissipative processes into the system's dynamics, the linear entropy behaves as a witness and no longer quantifies the entanglement.

The linear entropy is defined as follows,
\begin{equation}\label{Entropia-def}
\delta \left (\rho\right ) \equiv 1- \text{Tr}_2 \left (\rho_2^2\right ),
\end{equation}
where $\text{Tr}_i$  is the partial trace over the subsystem $i$ and $\rho_2$ is the reduced density operator, which is the result of tracing out the subsystem 1 from the system's density operator; that is,
\begin{equation}
\rho_2 = \text{Tr}_1 \left (\rho\right ).
\end{equation} 

\subsection{The Wigner representation}

The Wigner representation allows us to establish qualitatively whether a state is quantum or not. The Wigner representation in the phase space is positive-definite for classical states, so that negative values of the Wigner function are indicators of quantumness of the states. Furthermore, if the state under consideration in a superposition of other states, then the Wigner representation allows to observe the interference between those states, and if any of the state's quadrature is squeezed  \cite{Davidovich-1997, Gerry-book, Barnett-book}.

The Wigner representation is defined as,
\begin{equation}
W\left (\alpha \right )=2\, Tr \left [ D^{-1}\left (\alpha\right )\rho D \left (\alpha\right ) \mathcal{P} \right ],
\end{equation}
where $D\left (\alpha\right )$ is the displacement operator, $\alpha = x + i p$ with $x$ and $p$ the phase space axis, $\rho$ is the density operator describing the state under study, and $\mathcal{P}= \exp \left (i \pi a^{\dagger}a \right )$ is the parity operator.

\section{Results and Discussion}\label{sec:03}

\begin{figure*}
\begin{minipage}{0.46\textwidth}
\includegraphics[width=0.98\textwidth]{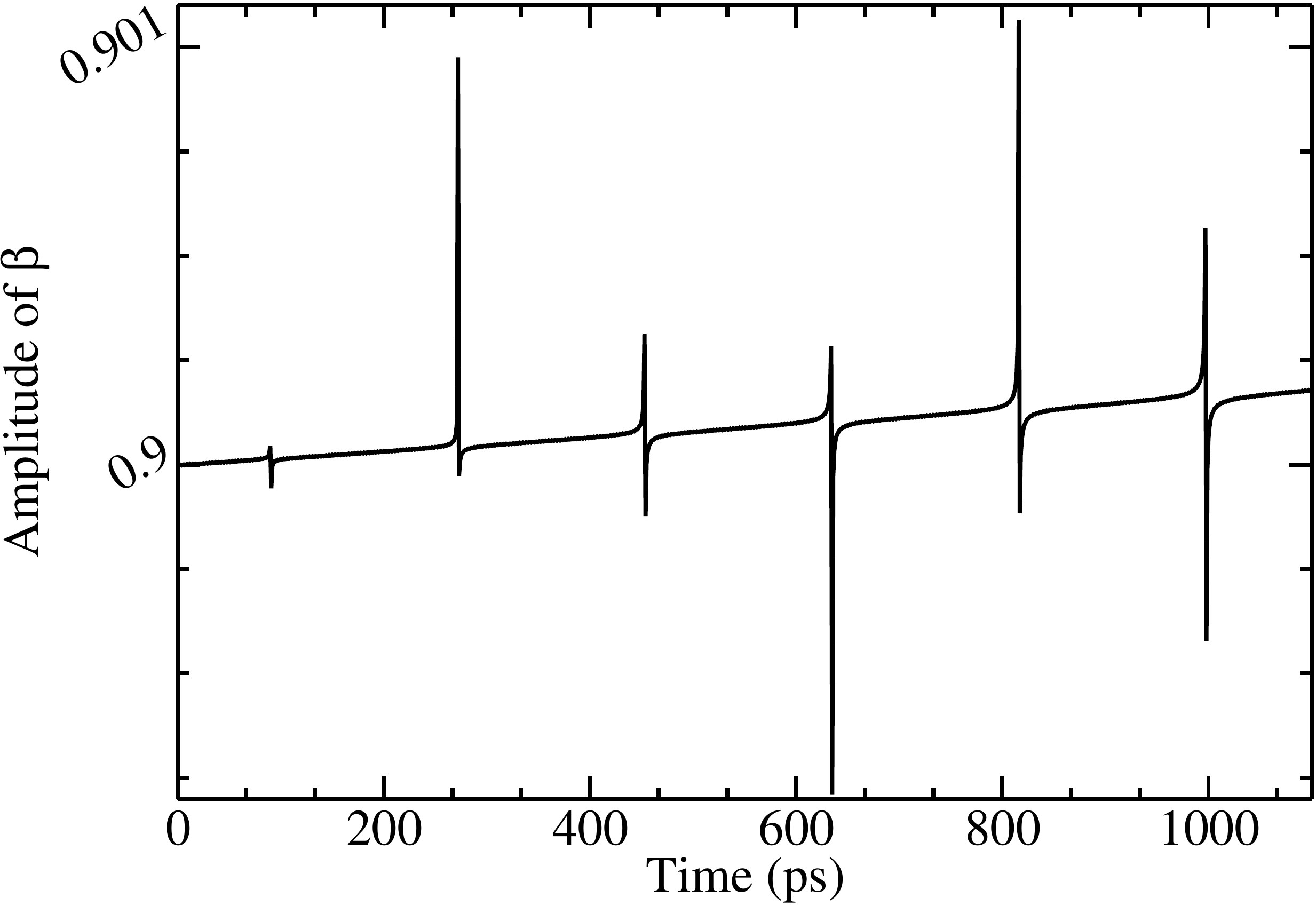} 
\end{minipage}
 \begin{minipage}{0.17\textwidth}
    \includegraphics[width=0.98\textwidth]{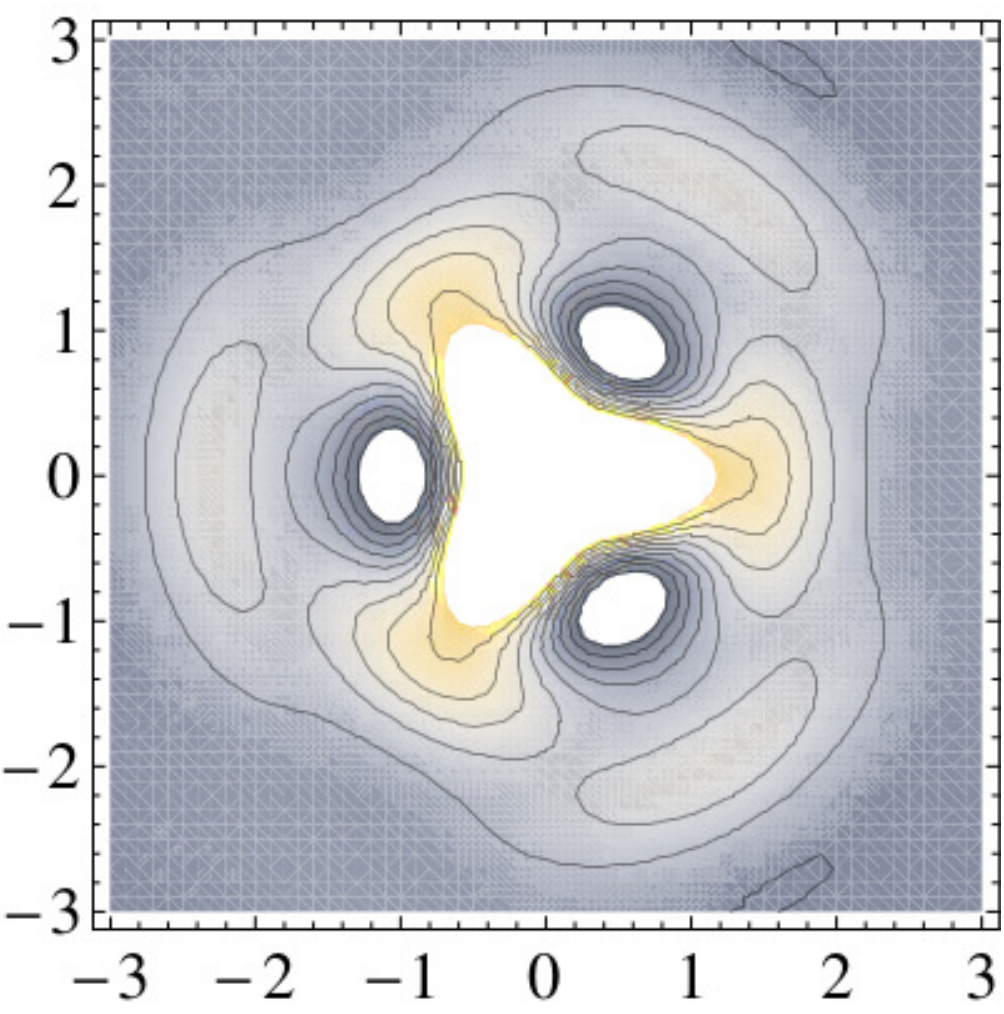}\\
    \includegraphics[width=0.98\textwidth]{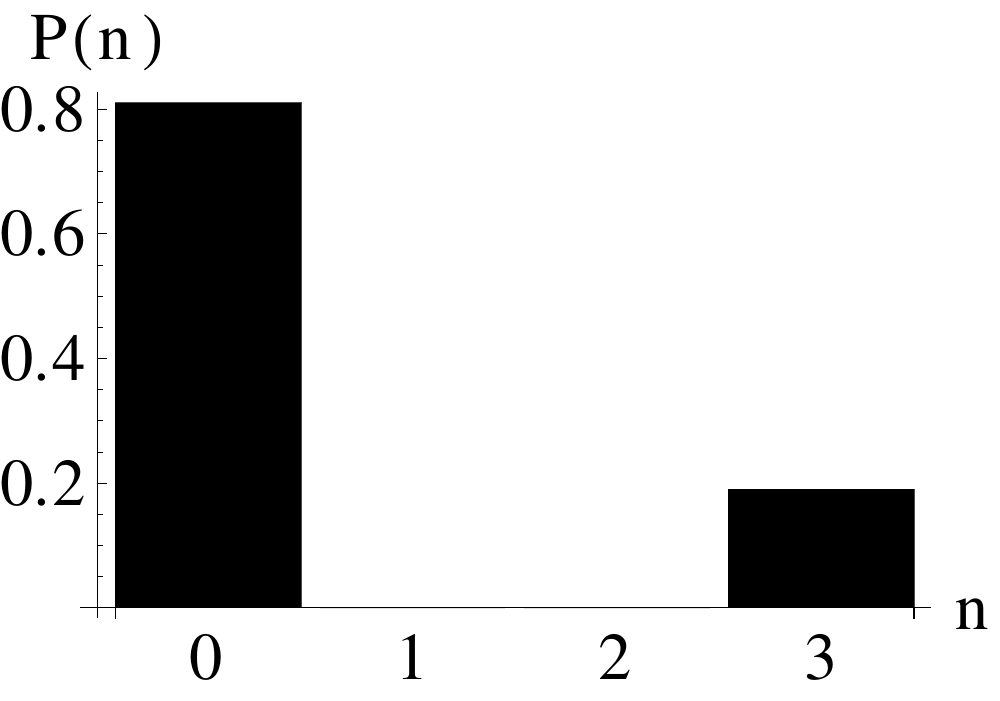}
    \begin{small}
    \begin{center}
    $t=0\,ps$
    \end{center}
    \end{small}
  \end{minipage}
  \ \hfill 
  \begin{minipage}{0.17\textwidth}
    \includegraphics[width=0.98\textwidth]{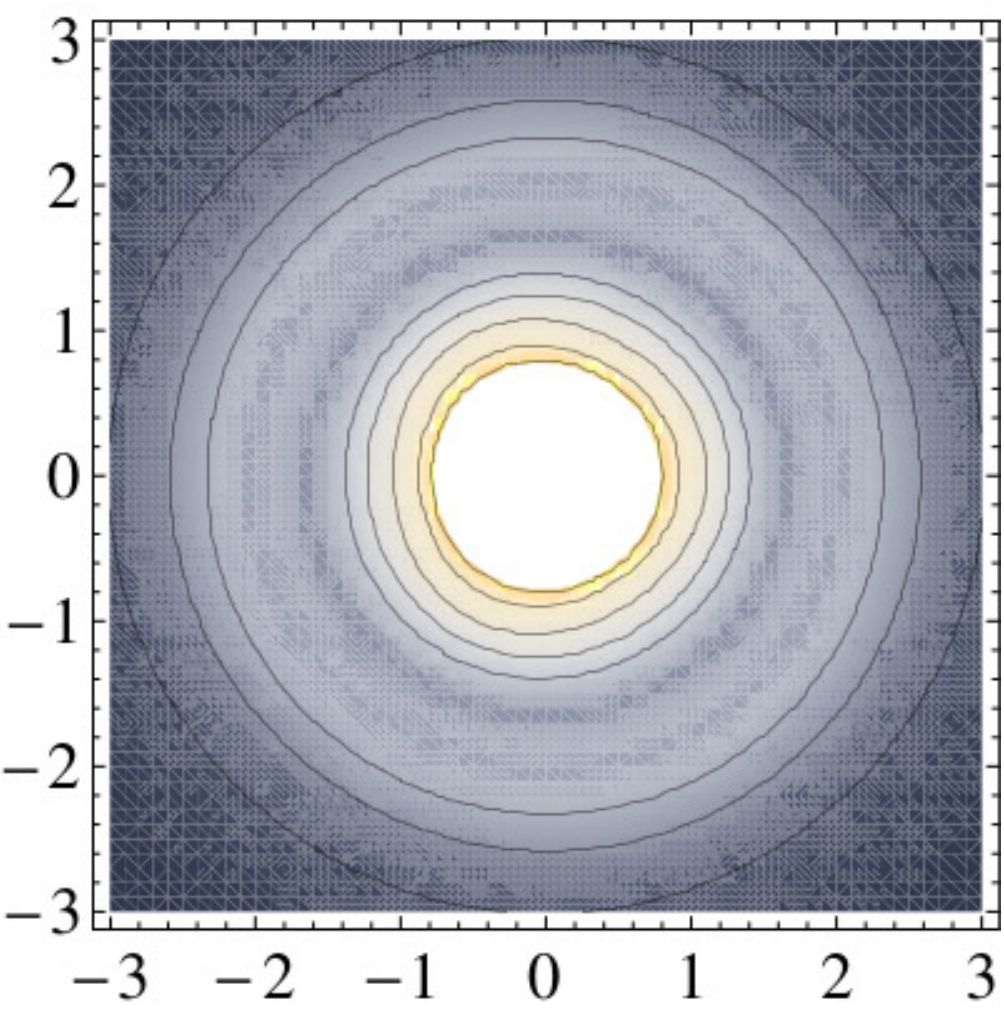}\\
    \includegraphics[width=0.98\textwidth]{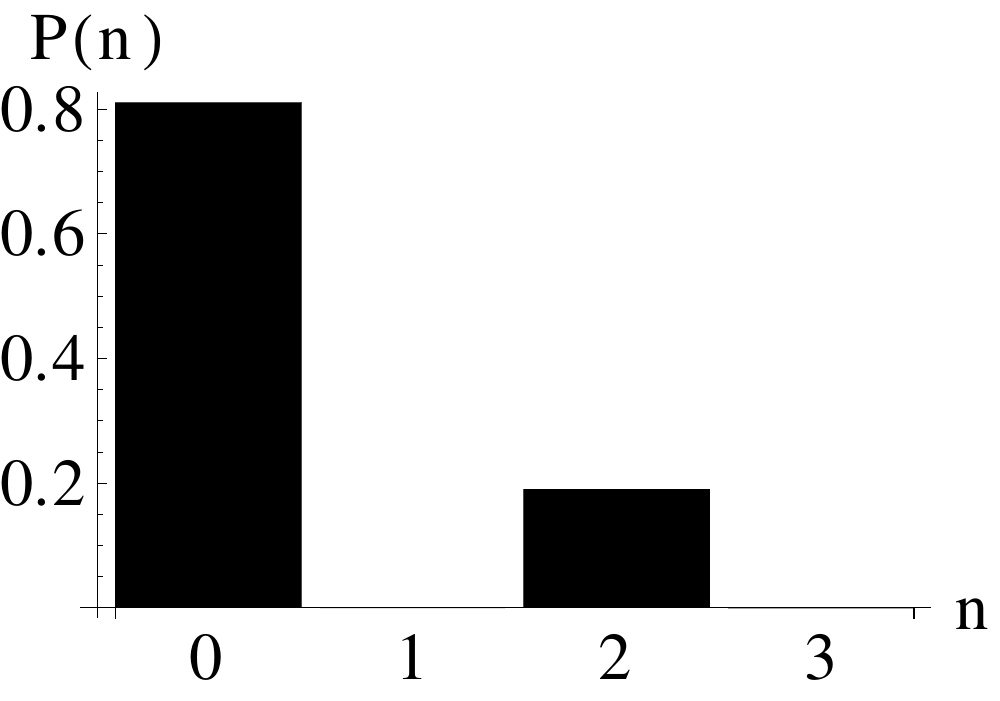}
    \begin{small}
    \begin{center}
    $t=90\,ps$
    \end{center}
    \end{small}
  \end{minipage}
  \ \hfill
  \begin{minipage}{0.17\textwidth}
    \includegraphics[width=0.98\textwidth]{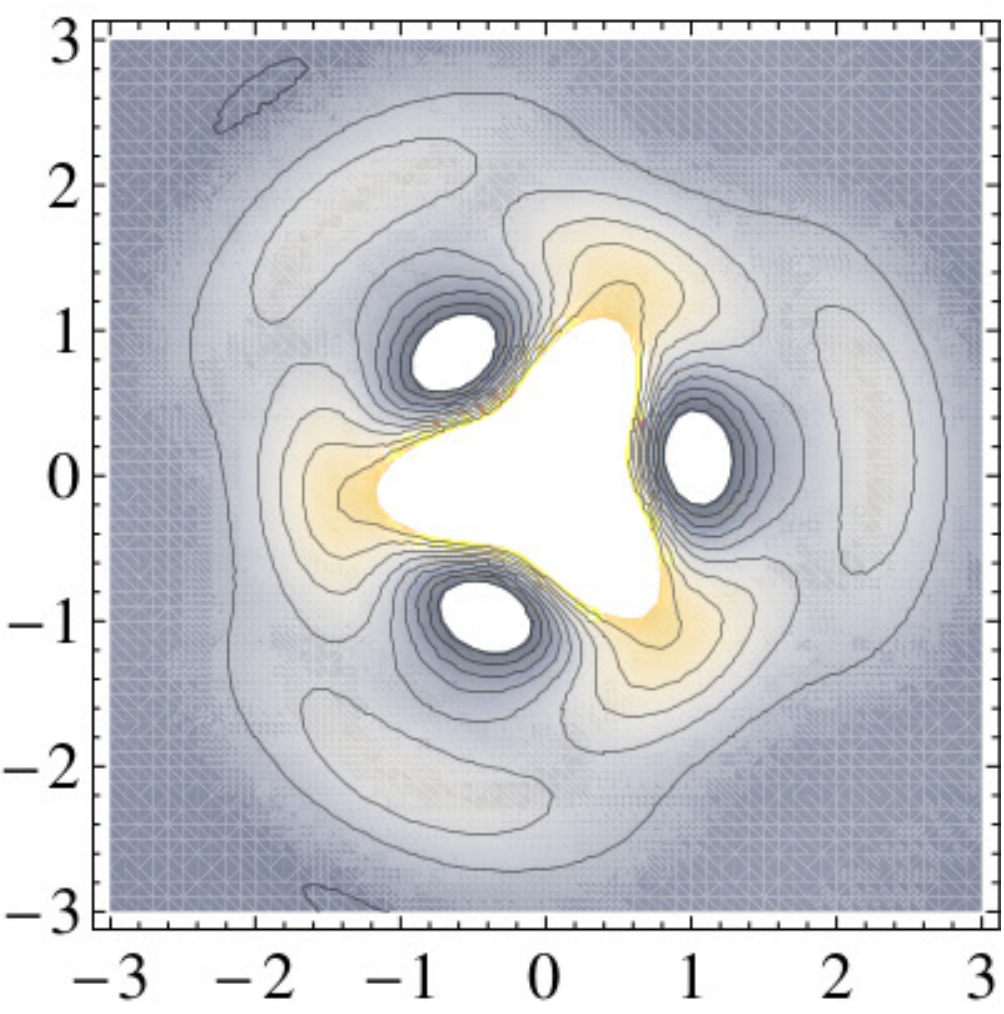}\\
    \includegraphics[width=0.98\textwidth]{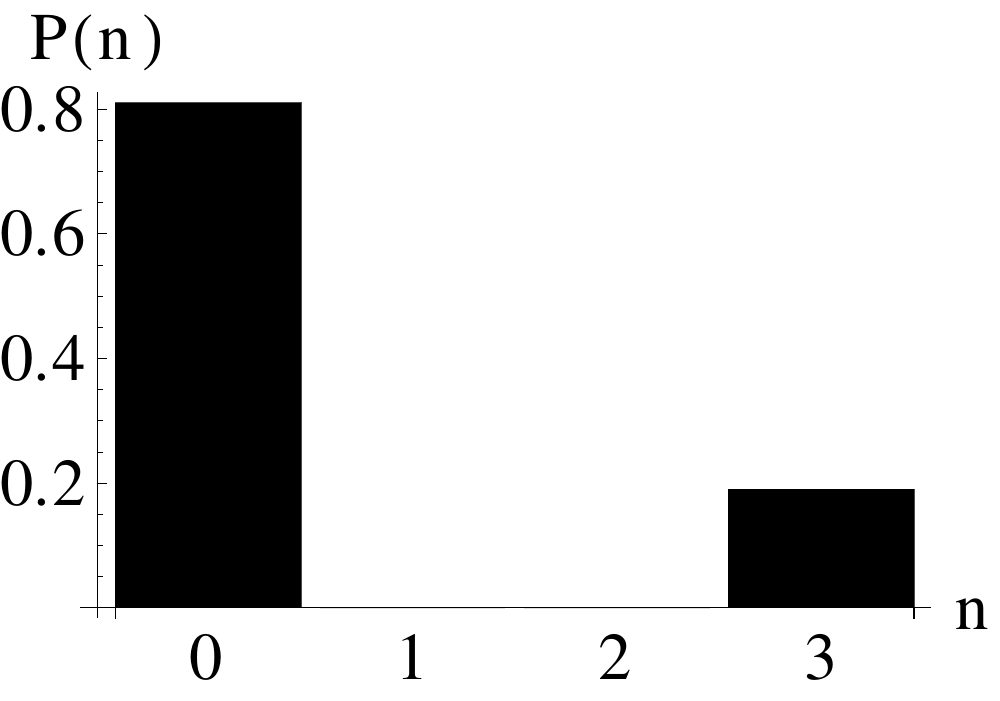}
    \begin{small}
    \begin{center}
    $t=180\,ps$
    \end{center}
    \end{small}
  \end{minipage}
  
\caption{\label{fig:beta(t)-base} In the left panel we present the temporal evolution of the parameter $\beta$ for the quantum dot initially in its ground state. A series of resonances may be seen, which are equally apart in time, and coincide with the time intervals in which the light state is the one given by eq. (\ref{Estella}). In the right panel we present the Wigner function and the population of the light's density operator, which verify the periodicity of the system's temporal evolution. In this case, the dipole coupling energy is $g=10$ ps$^{-1}$, and the temporal period is approximately $180$ ps.}
\end{figure*}

\begin{figure*}
\begin{minipage}{0.46\textwidth}
\includegraphics[width=0.98\textwidth]{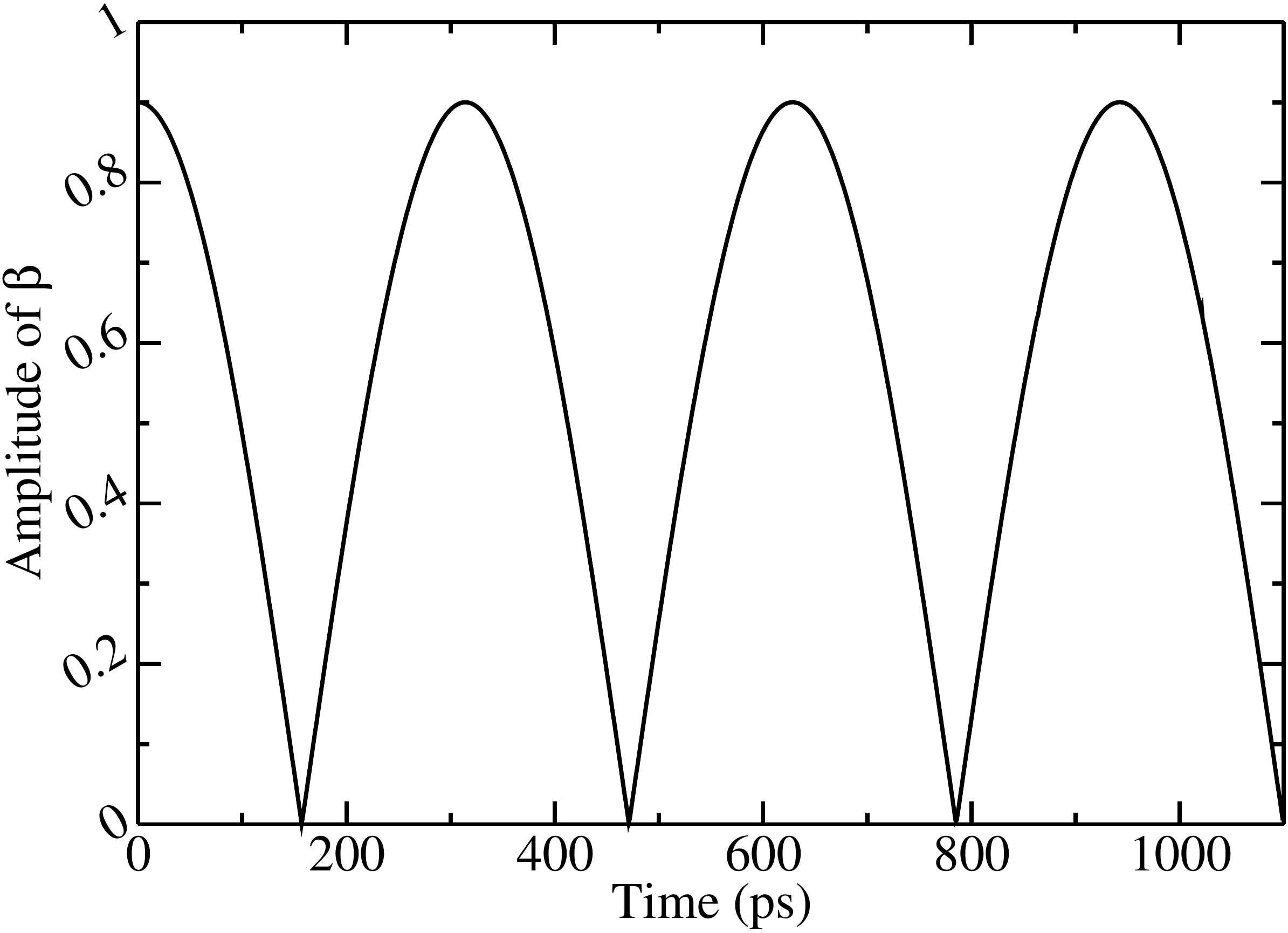} 
\end{minipage}
 \begin{minipage}{0.17\textwidth}
    \includegraphics[width=0.98\textwidth]{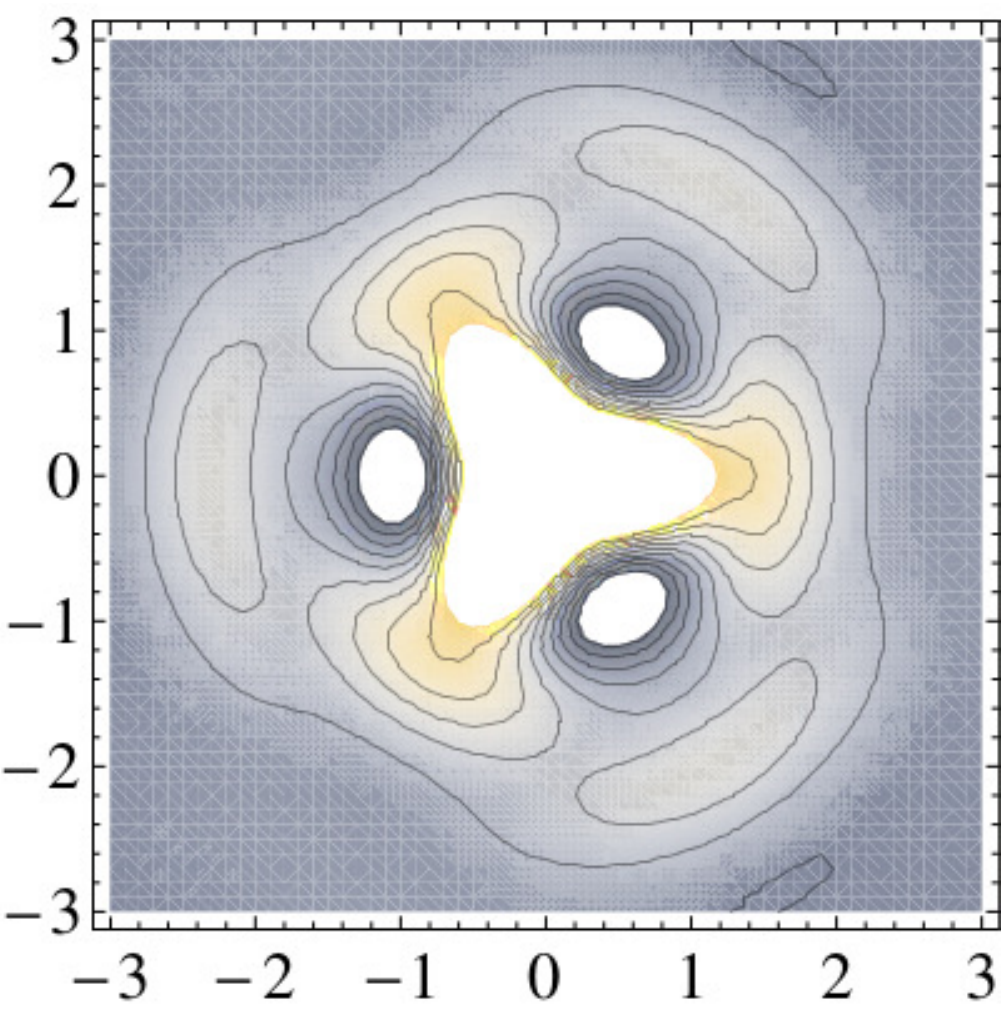}\\
    \includegraphics[width=0.98\textwidth]{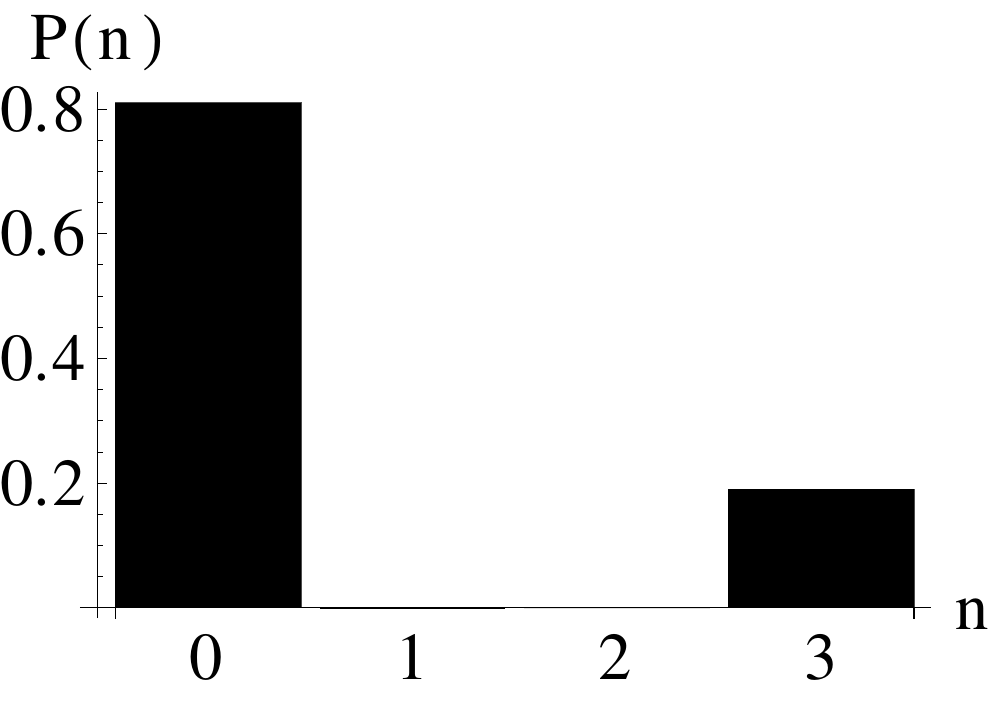}
    \begin{small}
    \begin{center}
    $t=0$ ps
    \end{center}
    \end{small}
  \end{minipage}
  \ \hfill 
  \begin{minipage}{0.17\textwidth}
    \includegraphics[width=0.98\textwidth]{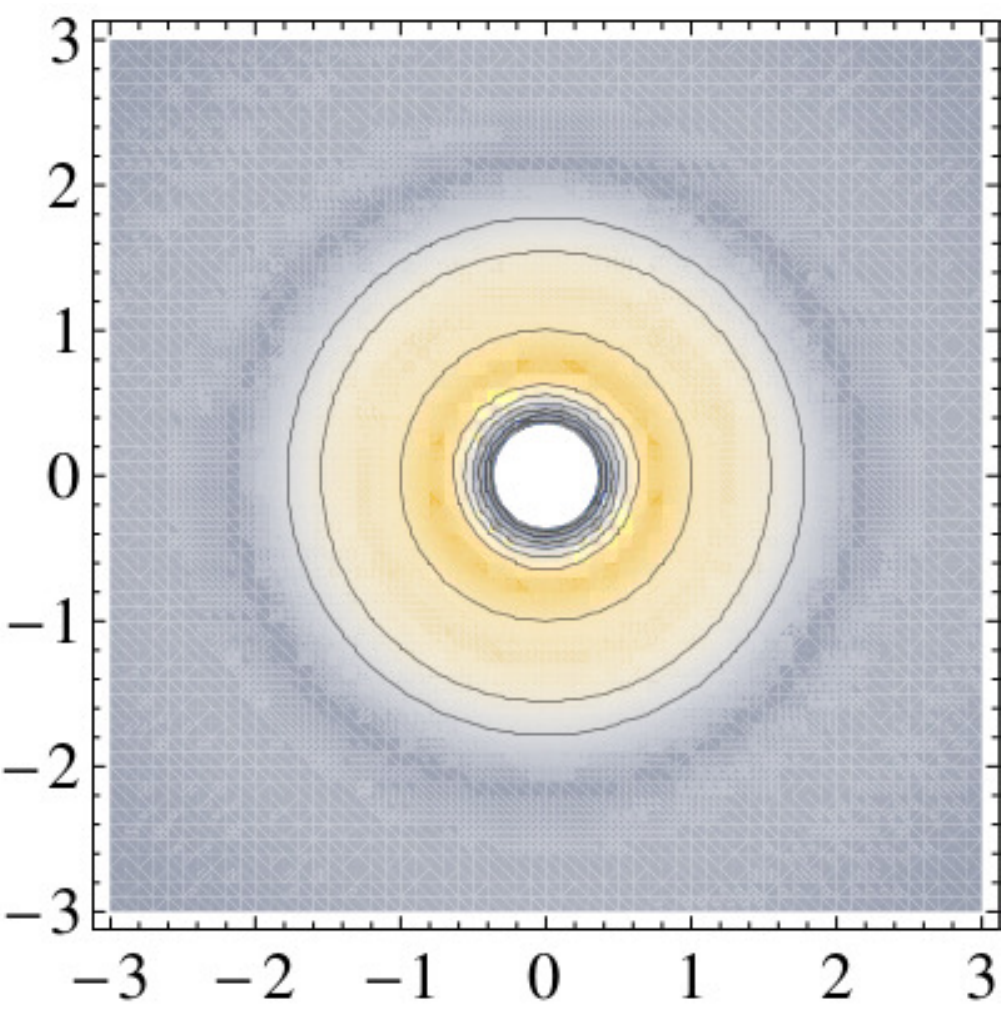}\\
    \includegraphics[width=0.98\textwidth]{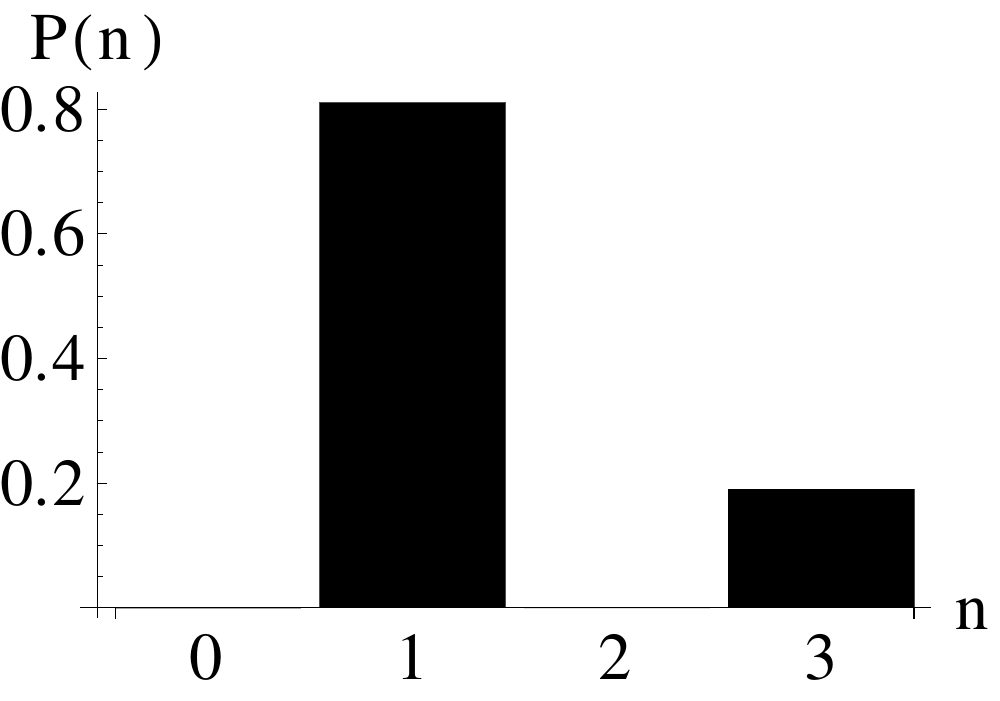}
    \begin{small}
    \begin{center}
    $t=156$.$6$ ps
    \end{center}
    \end{small}
  \end{minipage}
  \ \hfill
  \begin{minipage}{0.17\textwidth}
    \includegraphics[width=0.98\textwidth]{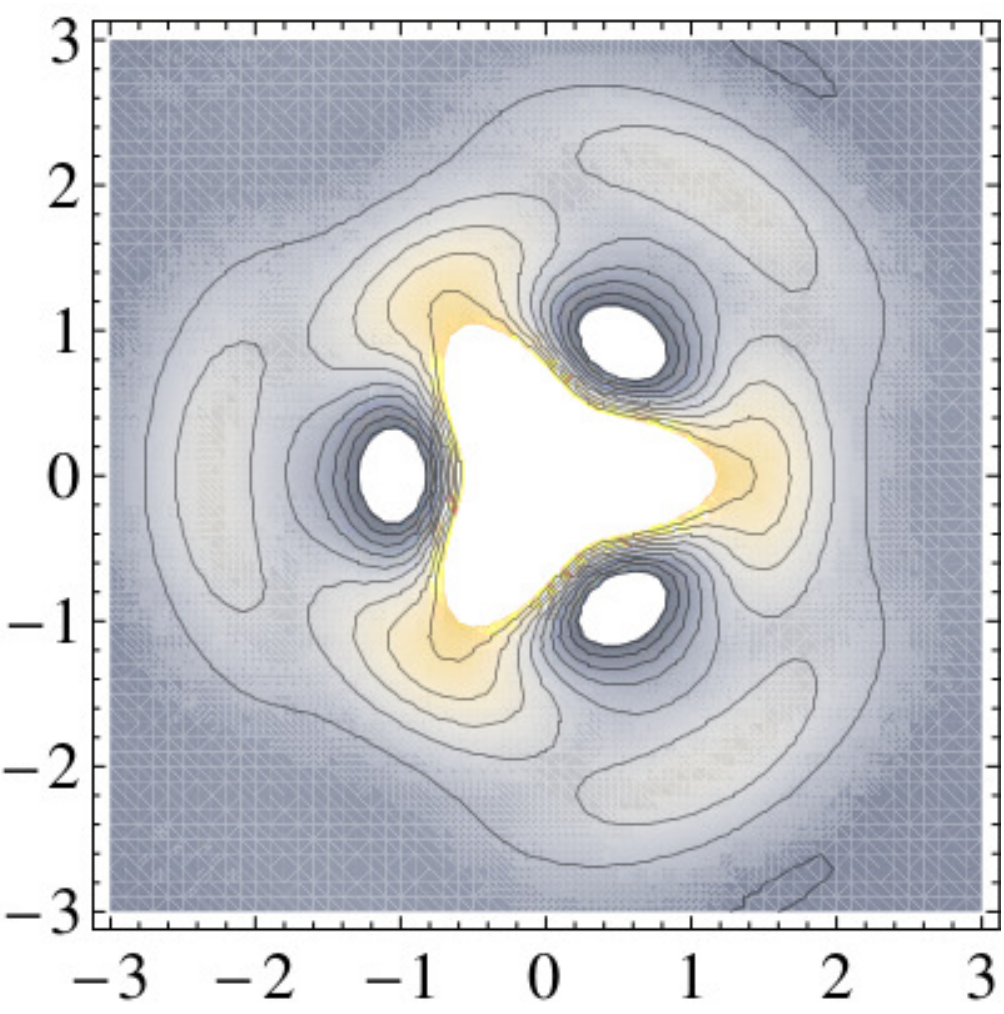}\\
    \includegraphics[width=0.98\textwidth]{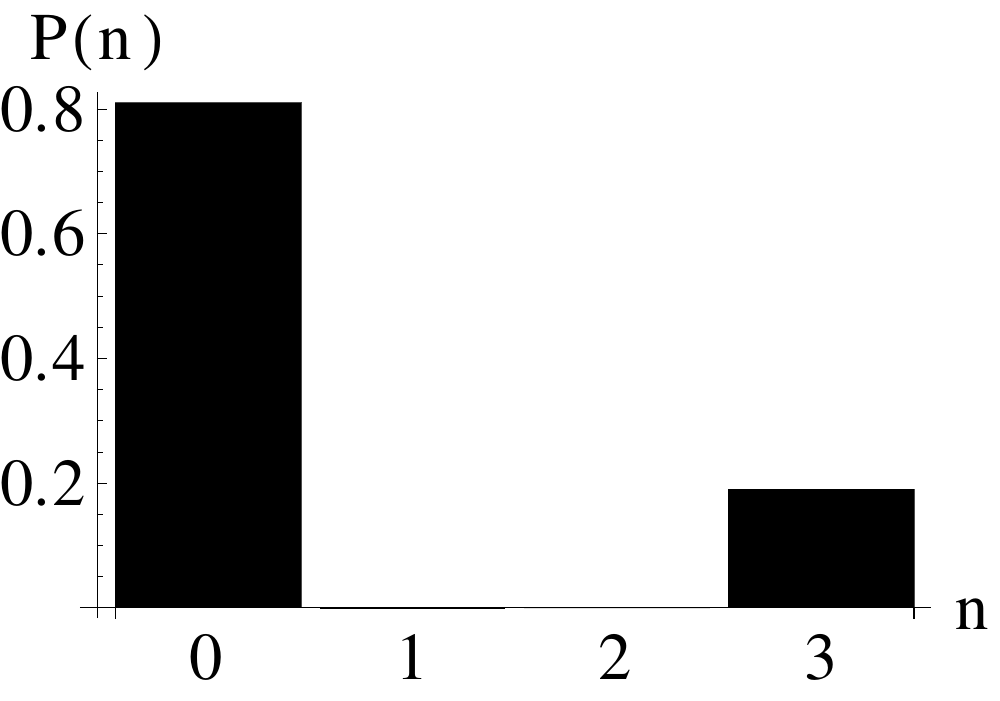}
    \begin{small}
    \begin{center}
    $t=313$.$1$ ps
    \end{center}
    \end{small}
  \end{minipage}
  
\caption{\label{fig:beta(t)-excitado} In the left panel we present the temporal evolution of the parameter $\beta$ for the quantum dot initially in its excited state. It is clear that the system has a periodic behavior. In the right panel we present the Wigner function and the population of the light's density operator, which verify the periodicity of the system's temporal evolution. In this case, the dipole coupling energy is $g=10$ ps$^{-1}$, and the temporal period is approximately $313$.$1$ ps.}

\end{figure*}

\begin{figure*}
\begin{minipage}{0.47\textwidth}
\centering \includegraphics[width=0.98\textwidth]{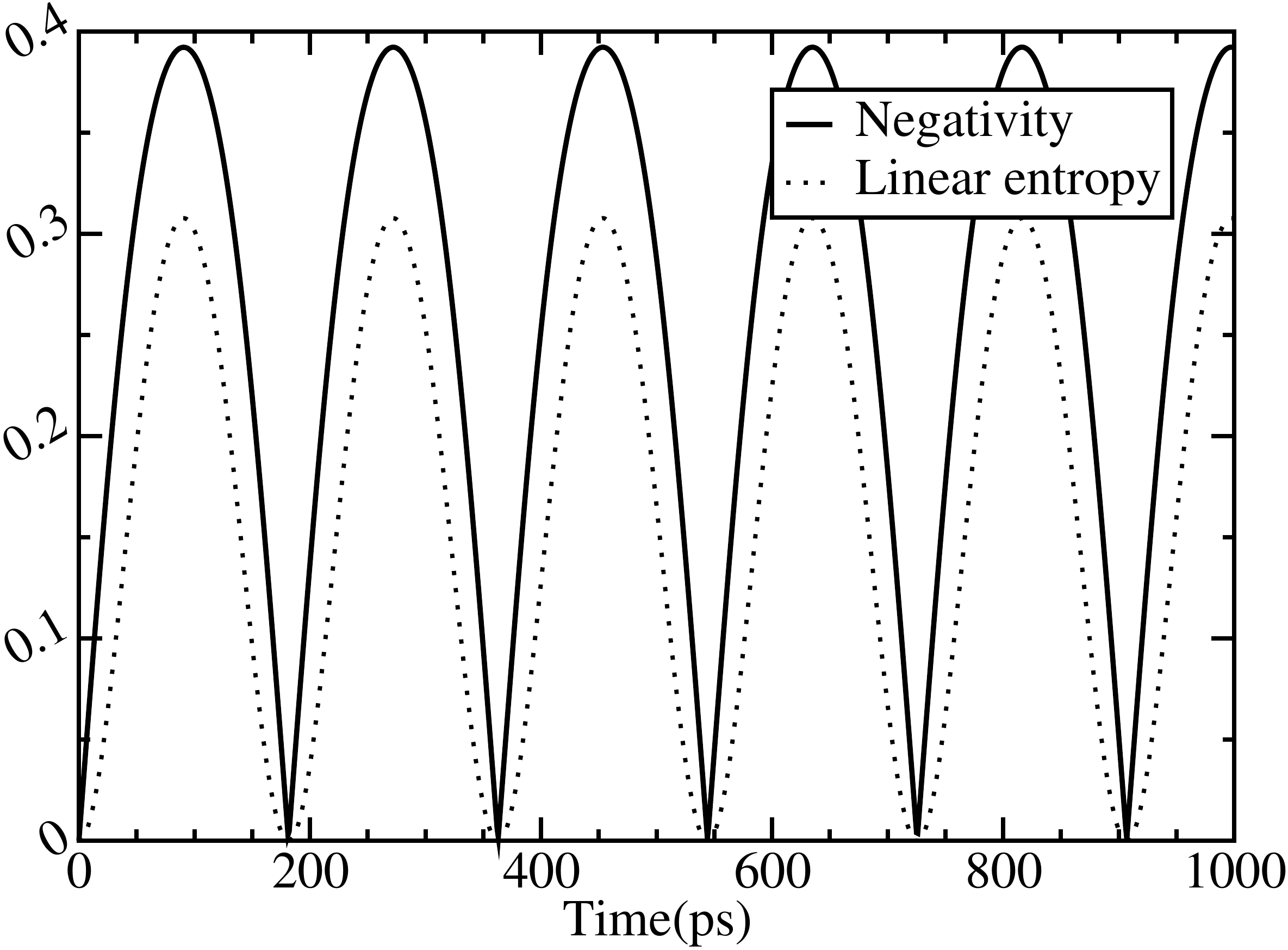} 
\end{minipage}
\ \hfill
 \begin{minipage}{0.47\textwidth}
\centering \includegraphics[width=0.98\textwidth]{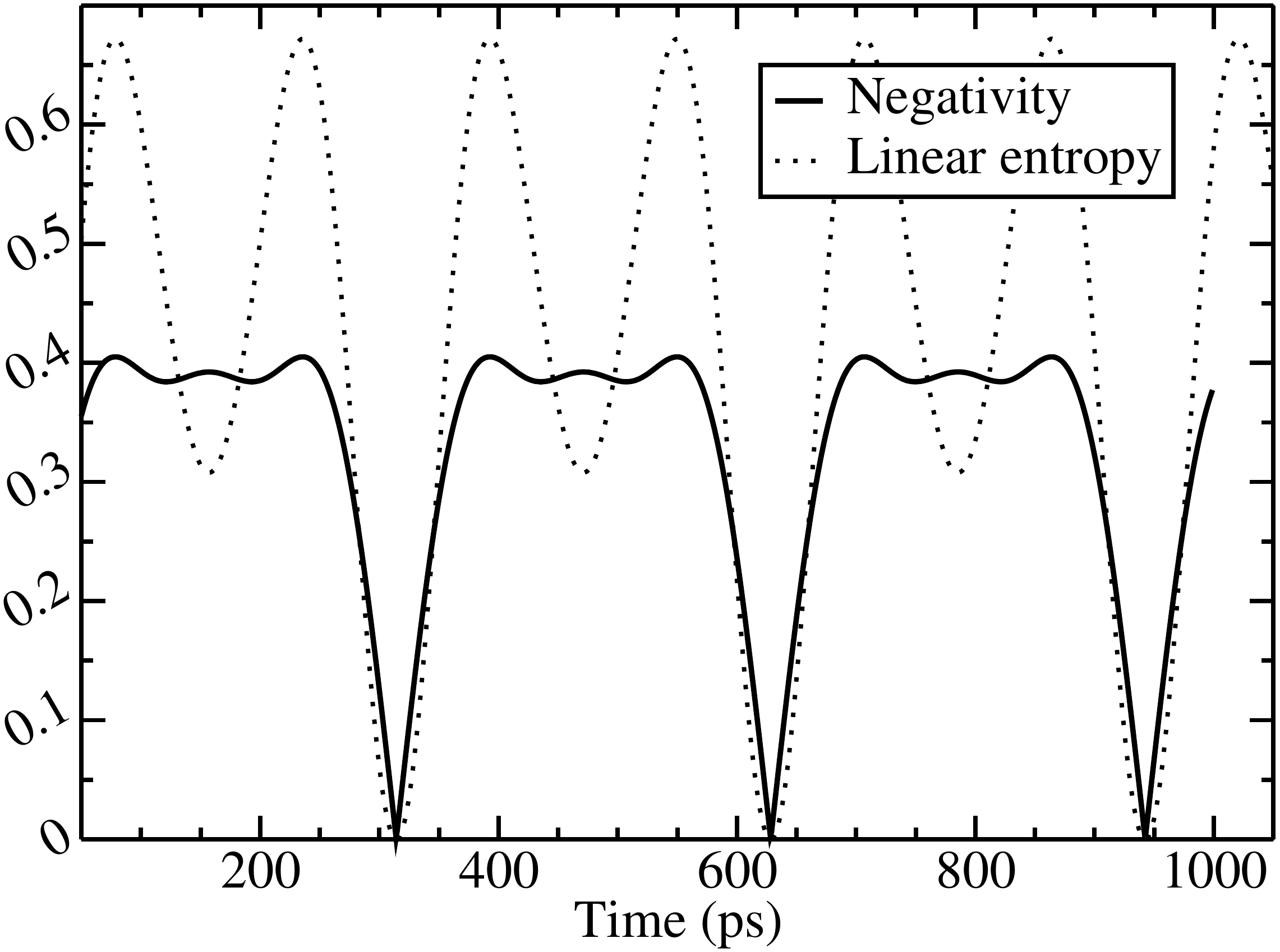}
  \end{minipage}
\caption[Negativity and linear entropy: ideal cavity]{Negativity and linear entropy of the system, considering the state of light with the initial condition given by eq. (\ref{cond.ini.}) with a)  $\theta=0$ (left)  and b) $\theta=\pi/2$ (right). It is straightforward to notice that the two functions have periodic behavior, and that its period is equal to the one of the  parameter $\beta$.}
\label{fig:Entrela-Hamilton}
\end{figure*}

\begin{figure*}
\begin{minipage}{0.47\textwidth}
\centering \includegraphics[width=0.98\textwidth]{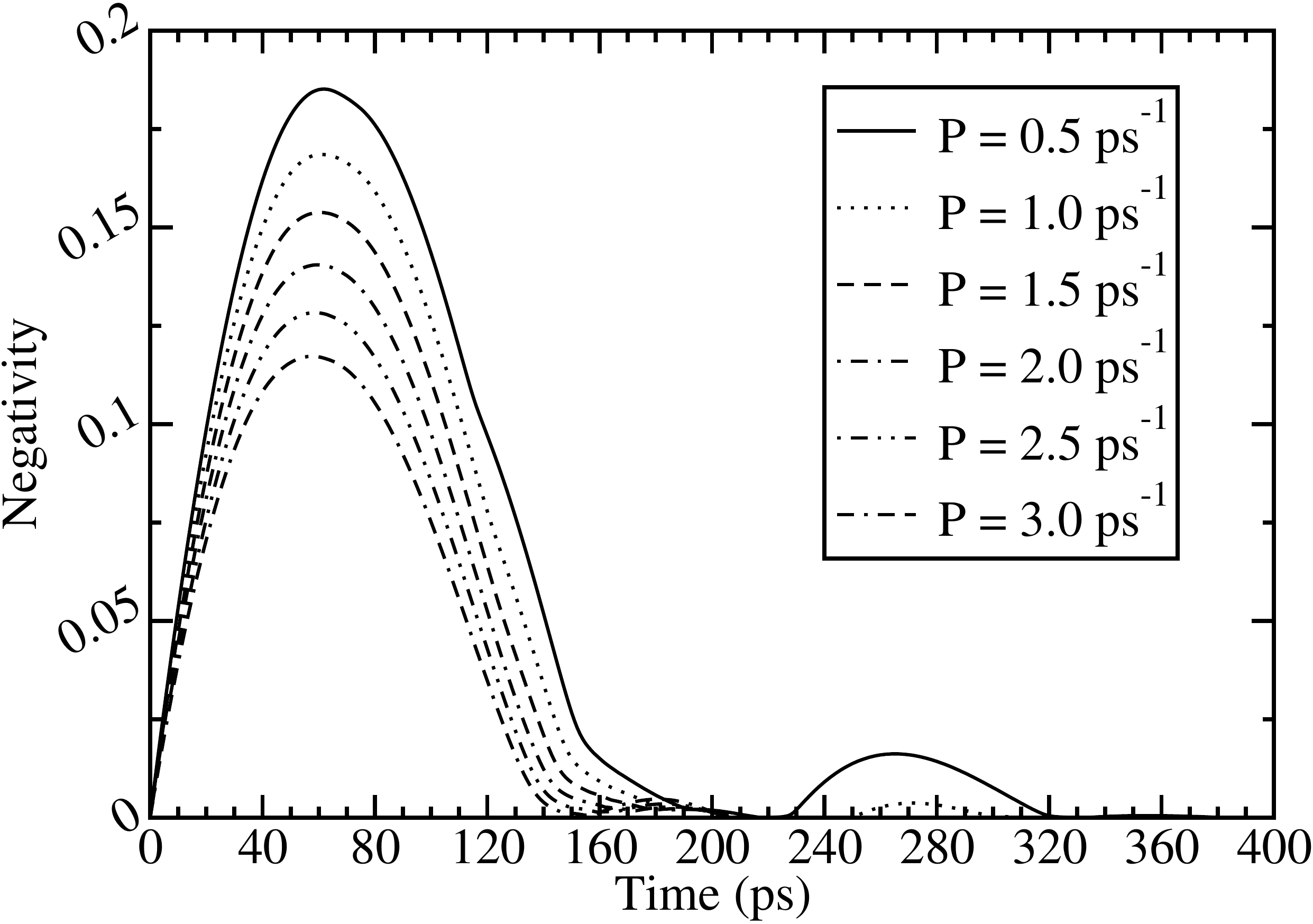} 
\end{minipage}
\ \hfill
 \begin{minipage}{0.47\textwidth}
\centering \includegraphics[width=0.98\textwidth]{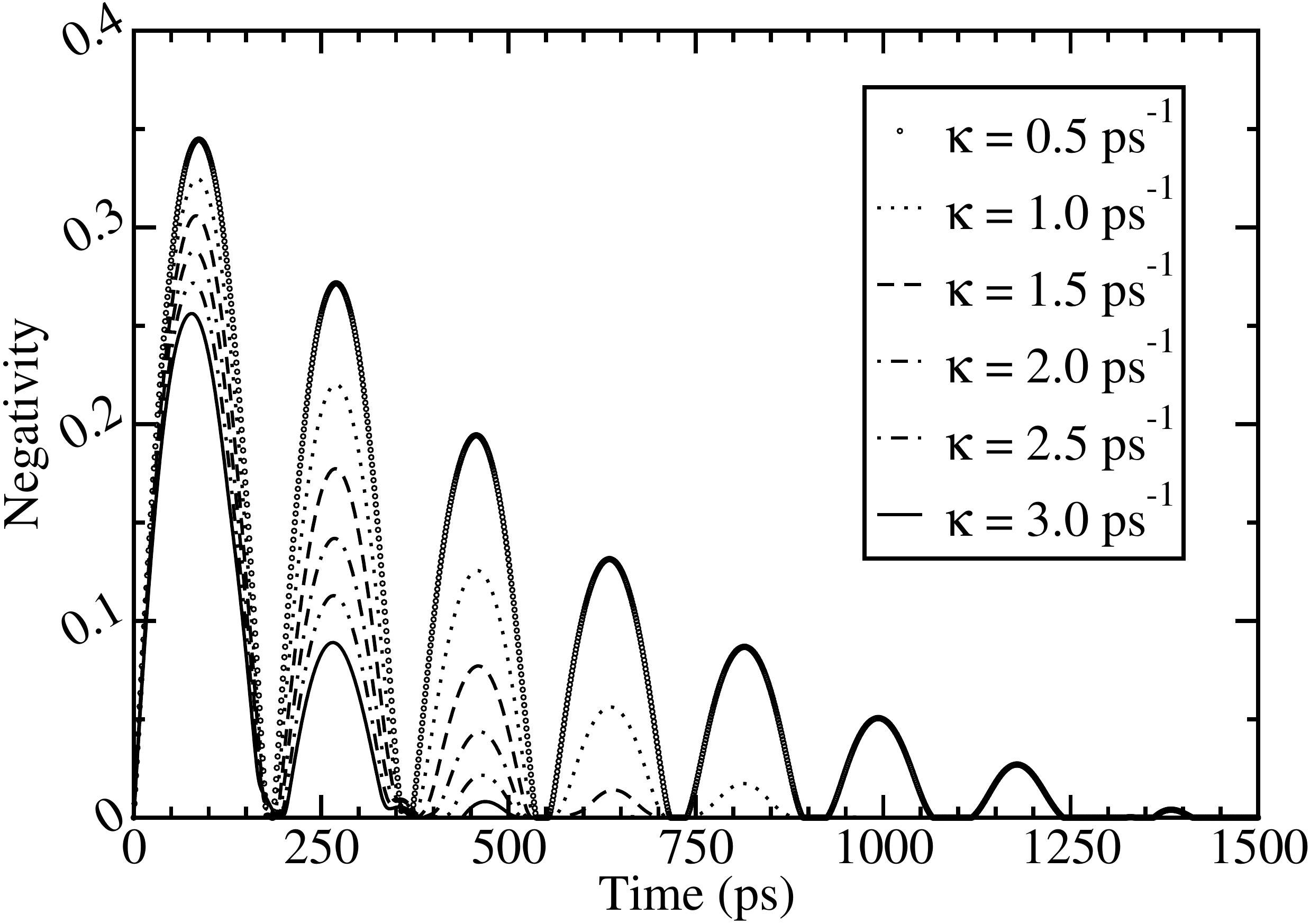}
  \end{minipage}
\caption[Negativity and linear entropy: real cavity (fixed photon leakage rate)]{
In the left panel we observe the negativity with a fixed photon leakage rate $\kappa=6$ ps$^{-1}$, for several excitation pumping rates. We observe that although increasing the excitation pumping rate makes the system get maximally entangled faster, the entanglement maximum is lower. Furthermore, in the right panel we present the negativity with a fixed excitation pumping rate $P=0$.$5$ ps$^{-1}$ and several photon leakage rates. We observe a series of damped oscillations between entangled and separable states, noticing that the damping rate increases as the $\kappa$ becomes larger.  }
\label{fig:Entrela-dis-k}
\end{figure*}

\subsection{Election of the parameters}

Our model has several parameters, so its paramount to establish the range of values in which we are going to do the study. The coupling energy between the quantum dot and the electromagnetic field is usually of the order of the meV, so we take $g= 10$ ps$^{-1}$. On the other hand, in systems of semiconductor cavities the energies of both the quantum dot and the electromagnetic field are of the order of  $1$ eV. Therefore we take $\omega_a=1$eV y  $ \omega_\sigma= (\omega-\Delta)$, where $\Delta$ is known as the detuning, and is usually or the order of the meV. Since this detuning is small compared to the transition energies, in this paper we restrict ourselves to the perfect resonance; i.e. $\Delta=0$.
Moreover, since we are studying the system under the low-excitation regime, the rate of incoherent excitation  pumping  $P$ varied between zero and $3$ ps$^{-1}$, whereas the cavity losses $\kappa$ ranged between zero and $6$ ps$^{-1}$.
Finally, since in \cite{Lopez-2014} was shown that the three photon state, as the one in eq. (\ref{Estella})  may be prepared inside the cavity, we take the parameter $\beta=0.9$, and set the initial condition of the system as,
\begin{equation}\label{cond.ini.}
\ket{\psi_{t=0}}=\left (\cos\theta\ket{g}+\sin\theta\ket{e}\right )\ket{*},
\end{equation}
where $\ket{g}(\ket{e})$ is the ground (excited) state of the two-level system, and $\ket{*}$ is the three photon state given in eq. (\ref{Estella}).

\subsection{Ideal cavity}

In order to establish the characteristics of the dynamics due to the coherent evolution  (i.e. due to the Hamiltonian) and those due to the incoherent evolution (i.e. due to the dissipative processes), we first consider an ideal cavity in which the system is note perturbed in any form by the environment. 
On the other hand, since the state of light of our system is fully characterized by the parameter $\beta$, we study its temporal evolution.

In the left panel of fig. (\ref{fig:beta(t)-base}) we present the behavior of $\beta$ as a function of time, corresponding to an initial state given by eq. (\ref{cond.ini.}) and with $\theta=0$. The most relevant feature of that figure is the set of resonances, all of which are equally separated in time. 
On the other side, in the right panel of fig. (\ref{fig:beta(t)-base}), we present the Wigner function as well as the populations of the state of light, noticing the same periodicity. Thus, it is clear that th resonances shown by the $\beta$ parameter coincide with states of light for which the Fock state with $n=3$ is not probable. 

Analogously, in the left panel of  the figure (\ref{fig:beta(t)-excitado}) we show the behavior of $\beta$ as a function of time, corresponding to an initial state as in eq. (\ref{cond.ini.}) and with $\theta=\pi/2$. For this initial condition, the periodic behavior of the light states is quite evident. In the right panel of fig. (\ref{fig:beta(t)-excitado}) we present the Wigner function as well as the populations of the state of light, noticing that the parameter $\beta$ reaches its minimum when the probability of having a vacuum Fock state is equal to zero. Thus, the periodicity of $\beta$ is associated to the periodicity of the state of light.

In this way, for a dipole coupling energy of $g=10$ ps$^{-1}$, the the period of the amplitude of the parameter $\beta$ is: a) $T \approx 180$ ps for the quantum dot initially in its ground state, and b) $T \approx 313$.$1$ ps for the quantum dot initially in its excited state. The values of the periods can be checked in the right panels of the fig. (\ref{fig:beta(t)-base}) and fig. (\ref{fig:beta(t)-excitado}), respectively.

On the other side, we are interested in the degree of  entanglement between the electromagnetic field and the quantum dot. We investigate such quantity using the linear entropy and the negativity. In the figure (\ref{fig:Entrela-Hamilton}) we show the entanglement dynamics considering the system's initial condition given by eq. (\ref{cond.ini.}), with  $\theta=0$ (left panel) and $\theta=\pi/2$ (right panel). 
In the first place, we notice that the negativity maxima coincide with the time intervals for which the state of light is a three-photon state; and that an analogous behavior is shown by the linear entropy. Nevertheless, for the quantum dot initially in its excited state, the linear entropy has a local minimum  in the middle of two consecutive negativity minima. The linear entropy's local minimum is associated to a zero probability of having the Fock state with $n=0$ in the state of light.

\subsection{Real cavity: dissipative dynamics}

The exposure to the environment and the external manipulation of the system produces the destruction of some of its properties, for example the periodical behavior. Moreover, using a specific set of parameters the system reaches an stationary state, which could be useful to recover the initial state and thus information about the original system.

Since the dynamical evolution is no longer expected to be periodic, the study of the parameter $\beta$ as a function of time is not as important as in the previous section. Nonetheless, the entanglement witnesses contain interesting information about the physical systems. Therefore, in this section we focus on the dynamics of those quantities as a function of the incoherent excitation pumping rate and photon leakage rate. 

In the left panel of figure (\ref{fig:Entrela-dis-k}) we present the behavior of the negativity as a function of the excitation pumping rate, keeping the photon leakage rate constant at $\kappa=6$ ps$^{-1}$. In the first place, we observe that the negativity reach an stationary value, which implies that the observed behavior is independent of the system's initial state. On the other side, we notice that by increasing the values of $P$ the negativity reaches its maximum faster but the value of the maximum is also reduced. 

The study of the dynamical behavior of the entanglement witnesses as a function of the photon leakage rate is shown in the right panel of figure (\ref{fig:Entrela-dis-k}). We notice that at certain time intervals the negativity becomes zero, and then takes positive values again. This phenomena is usually refer to as \textit{entanglement sudden death} and has been observed in Jaynes-Cummings-like systems \cite{Eberly-2006, Eberly-2004, Eberly1-2006}. Furthermore, we observe that both the value at the negativity maxima and the number of negativity \textit{revivals} diminishes as the photon leakage rate $\kappa$ increases.

\section{Conclusions}\label{sec:04}

In this paper we study the dynamical behavior of a quantum state of light, known as a three-photon state, interacting with a quantum dot inside a cavity. In the first place we analyze the temporal evolution of the system inside a perfect cavity; i.e. neglecting all incoherent processes. Thus, we observe a periodic evolution, which depends strongly on the quantum dot's initial condition. Likewise, the negativity and linear entropy have a periodic behavior, which period matches the one of the system's evolution.

Furthermore, we investigate the temporal evolution of the system considering a cavity with dissipative processes; such as an incoherent excitation pumping rate and the leakage of photons out of the cavity. Keeping the excitation pumping rate constant, we observe time intervals for which the entanglement vanishes.

Finally, we report that although the system is quite interesting because it can readily be prepared in a semiconductor microcavity, the environment destroys the entanglement between the three photon state and the quantum dot in a time scale too small for the system to be useful for quantum information processing.


\begin{acknowledgments}

This research has been supported by the Direcci\'{o}n de Investigaci\'{o}n - Sede Bogot\'{a}, Universidad Nacional de Colombia (DIB-UNAL) within the project No. 12584. 
Furthermore, we acknowledge technical support of the Grupo de \'{O}ptica e Informaci\'{o}n Cu\'{a}ntica.
\end{acknowledgments}



\begin{thebibliography}{99}


\bibitem{Tejedor-2004}  J.I. Perea, D. Porras and C. Tejedor. \textit{Phys. Rev. B}, {\bf 70}, 2004, p. 115304.

\bibitem{Vinck-2005}  C.A. Vera, N. Quesada and H. Vinck-Posada. \textit{J. Phys.: Condens. Matter}, {\bf 21}, 2005, p. 395603.

\bibitem{Yamamoto-2000} O. Benson, C. Santori, M. Pelton and Y. Yamamoto. \textit{Phys. Rev. Lett.}, {\bf 84}, 2000, p. 2513.

\bibitem{Haroche-2007} S. Gleyzes, S. Kuhr, C. Guerlin, J. Bernu, S. Del\'eglise, U. Busk Hoff, M. Brune, J.M. Raimond, and S. Haroche, \textit{Nature} \textbf{446}, 297 (2007).

\bibitem{Bundler-2014} C. S\'anchez Mu\~noz,	 E. del Valle,	 A. Gonz\'alez Tudela,	 K. M\"uller,	 S. Lichtmannecker,	 M. Kaniber, C. Tejedor,	 J. J. Finley and F. P. Laussy, \textit{Nature Photon.} Advance online publication. doi:10.1038/nphoton.2014.114

\bibitem{Yamamoto-2002} M. Pelton, C. Santori, B. Zhang, G. Solomon, J. Plant and Y. Yamamoto. \textit{Phys. Rev. Lett.}, {\bf 89}, 2002, p. 233602.

\bibitem{Yamamoto-2003}  J. Vuckovic, D. Fattal, C. Santori, G.S. Solomon and Y. Yamamoto. \textit{Appl. Phys. Lett.}, {\bf 82}, 2003, p. 3596.

\bibitem{Pelton-2003} G. Solomon, M. Pelton and Y. Yamamoto. \textit{Phys. Rev. Lett.}, {\bf 86}, 2003, p. 3903.


\bibitem{Kiesel-2003} N. Kiesel, M. Bourennane, C. Kurtsiefer, H. Weinfurter, D. Kaszlikowski, W. Laskowski and M. Zurowski. \textit{J. Mod. Opt.}, {\bf 50}, 2003, p. 1131.

\bibitem{Douady-2004} J. Douady and B. Boulanger. \textit{Opt. Lett.}, {\bf 29}, 2004, p. 2794.

\bibitem{Corona-2011} M. Corona, K. Garay-Palmett and A.B. U'Ren. \textit{Opt. Lett.}, {\bf 36}, 2011, p. 190.

\bibitem{Gravier-2008} F. Gravien and B. Boulanger. \textit{J. Opt. Soc. Am. B}, {\bf 25}, 2008, p. 98.

\bibitem{Richard-2011} S. Richard, K. Bencheikh, B. Boulanger and J.A. Levenson. \textit{Opt. Lett.}, {\bf 36}, 2011, p. 3000.

\bibitem{Dot-2012} A. Dot, A. Borne, B. Boulanger, K. Bencheikh and J.A. Levenson. \textit{Phys. Rev. A}, {\bf 85}, 2012, p. 023809.

\bibitem{Persson-2004} J. Persson, T. Aichele, V. Zwiller, L. Samuelson and O. Benson. \textit{Phys. Rev. B}, {\bf 69}, 2004, p. 233314.

\bibitem{Antonosyan-2011} D.A. Antonosyan, T.V. Gevorgyan and G.Y. Kryuchkyan. \textit{Phys. Rev. A}, {\bf 83}, 2011, p. 043807.

\bibitem{Rodrigo-2011} S.G. Rodrigo, S. Carretero-Palacios, F.J. García-Vidal and L. Martín-Moreno. \textit{Phys. Rev. B}, {\bf 83}, 2011, p. 235425.

\bibitem{Lopez-2014} J.C. L\'opez Carre\~no and H. Vinck Posada, \textit{Phys. Scr.} \textbf{T160}, 014027 (2014).

\bibitem{May-2005} J.C. May, J.H. Lim, I. Biaggio, N.N.P. Moonen, T. Michinobu and F. Diederich \textit{Opt. Lett.}, {\bf 30}, 2005, p. 223057.

\bibitem{JCModel} E. Jaynes and F. Cummings, {\it Proc. IEEE} \textbf{51}, 81, 89 (1963).


\bibitem{Negat} G. Vidal and R.F. Werner, \textit{Phys. Rev. A} \textbf{65}, 032314 (2002).

\bibitem{Peres-1996}  A. Peres. \textit{Phys. Rev. Lett.}, {\bf 77}, 1996, p. 1413.

\bibitem{Horodecki-1996} M. Horodecki et al. \textit{Phys. Lett. A}, {\bf 223}, 1996, p. 1.

\bibitem{Farsi-2007} S.J. Akhtarshenas and M. Farsi. \textit{Phys. Scr.}, {\bf 75}, 2007, p. 608.

\bibitem{Faruya-1998} K. Faruya, M.C. Nemes and G.Q. Pellegrino. \textit{Phys. Rev. Lett.}, {\bf 80}, 1998, p. 5524.

\bibitem{Bose-2000} S. Bose and V. Vedral. \textit{Phys. Rev. A}, {\bf 61}, 2000, p. 040101(R).

\bibitem{Munro-2001} W.J. Munro, D.F.V. James, A.G. White and P.G. Kwiat. quant-ph/0103113v1, 2001.

%

\bibitem{Davidovich-1997}  L.G. Lutterbach and L. Davidovich. \textit{Phys. Rev. Lett.}, {\bf 78}, 1997, p. 2547.

\bibitem{Gerry-book}  C.C. Gerry and P.L. Knight, {\it Introductory Quantum Optics} (Cambridge: Cambridge University Press, 2004).

\bibitem{Barnett-book}  S.M. Barnett and P.M. Radmore, {\it Methods in Theoretical Quantum Optics} (Oxford Science Publications, 1997).

\bibitem{Eberly-2006} M. Yönac, T. Yu and J.H. Eberly, {\it J. Phys. B: At. Mol. Opt. Phys.}, {\bf 39}, 2006, p. S621.

\bibitem{Eberly-2004} T. Yu and J.H. Eberly, {\it Phys. Rev. Lett.}, {\bf 93}, 2004, p. 140404.

\bibitem{Eberly1-2006} T. Yu and J.H. Eberly, quant-ph/0602196v1, 2006.


\end{thebibliography}
 \end{document}